# Manifestation of charged and strained graphene layers in the Raman response of graphite intercalation compounds

J.C. Chacón-Torres

*Faculty of Physics, University of Vienna,*
*Strudlhofgasse 4, A-1090 Vienna, Austria*

L. Wirtz

*Physics and Material Sciences Research Unit,*
*Campus Limpertsberg,*
*University of Luxembourg,*
*L-1511 Luxembourg*
*and*
*Institute for Electronics, Microelectronics,*
*and Nanotechnology (IEMN),*
*CNRS-UMR 8520, Dept. ISEN,*
*59652 Villeneuve d'Ascq, France*

T. Pichler[*]

*Faculty of Physics, University of Vienna,*
*Strudlhofgasse 4, A-1090 Vienna, Austria*

## Abstract

We present detailed multi frequency resonant Raman measurements of potassium graphite intercalation compounds (GICs). From a well controlled and consecutive in-situ intercalation and high temperature de-intercalation approach the response of each stage up to *stage VI* is identified. The positions of the G and 2D lines as a function of staging depend on the charge transfer from K to the graphite layers and on the lattice expansion. Ab-initio calculations of the density and the electronic band-structure demonstrate that most (but not all) of the transferred charge remains on the graphene sheets adjacent to the intercalant layers. This leads to an electronic decoupling of these "outer" layers from the ones sandwiched between carbon layers and consequently to a decoupling of the corresponding Raman spectra. Thus higher stage GICs offer the possibility to measure the vibrations of single, double, and multi-layer graphene under conditions of biaxial strain. This strain can additionally be correlated to the in-plane lattice constants of GICs determined by x-ray diffraction. The outcome of this study demonstrates that Raman spectroscopy is a very powerful tool to identify local internal strain in pristine and weakly charged single and few-layer graphene and their composites, yielding even absolute lattice constants.

---

*thomas.pichler@univie.ac.at; `http://epm.univie.ac.at`

Raman spectroscopy is a key tool to probe the physical and electronic properties in graphene based materials [1–3]. Analysis of the two main signals in the Raman spectra, the so-called G-line around 1582 cm$^{-1}$ and the dispersive double-resonance peak in the range between 2600 and 2700 cm$^{-1}$ (which carries both the name G' line and 2D line in the literature), offers detailed information: e.g., it allows to determine the number of graphene layers, [2, 4] induced strain in the structure, [5–8] and charging [9–12]. We will show in this contribution how layer stacking, strain, and charging influence in detail the Raman spectra of graphite intercalation compounds.

Graphite intercalation compounds (GICs) consist of a consecutive stacking of graphene layers with intercalated alkali-metals, alkali-earth-metals, or rare-earth elements as well as p-type dopands like $FeCl_3$ or $AsF_5$ in between [13, 14]. GICs are classified in stages *I*, *II*, *III*, ..., where stage $n$ means that one intercalant layer follows after $n$ (usually AB stacked) graphene layers. The structural characterization of the different GICs is based on x-ray diffraction (XRD) and revealed a linear in-plane lattice expansion in GICs [15] as a function of inverse stage for *stage III* to *VI*. Later, a combination of Raman spectroscopy and XRD, [16, 17] was used to assign and analyze the vibronic structure for *stage I* to *stage VI* in Alkali GICs. Beyond *stage III*, one can distinguish between "outer" graphene layers that are adjacent to an intercalant layer and usually heavily charged and the "inner" layers that carry very little charge. We will refer in the following to charged (c) and uncharged (uc) layers. This *Nearest Layer* (NL) model was invoked in order to explain the splitting of the G line for Alkali GICs of *stage III* and higher [17]. The lower-frequency G-line at about 1580 cm$^{-1}$ corresponds in position to the G-line of neutral graphite and was therefore ascribed to the uc layers (supported by the fact that it is absent in stage I and stage II GICs. The higher-frequency line at about 1610 cm$^{-1}$ (which is present for all stages) was consequently ascribed to the c layers, even though the exact mechanism of the stiffening remained unclear at the time. The stiffening has by now been understood as the effect of non-adiabaticity onto the vibrations of charged graphene layers [18] and *ab-initio* calculations of *stage I* GICs have indeed confirmed the strong blue-shift of the G-line [19]. The NL model has been successful in identifying the different stages through the relative intensities of the G-lines of uc and c layers. However, the model alone is not able to explain the subtle frequency shifts of the G-line for the different stages and has not yet been used to investigate the dependence of the 2D line on staging.

In this article, we present an extensive Raman spectroscopy study of potassium GICs, measuring both the G and the 2D lines for different laser energies and for the different intercalation stages up to *stage VI*. The measurements are accompanied by *ab-initio* calculations of the electronic structure, charge transfer, lattice expansion and vibrational properties of these GICs. We present a quantitative refinement of the nearest layer model which takes into account the exact charge transfer, the lattice expansion and the effect of AB-stacking of the inner (uc) layers. By comparing with the available experimental data on charged, strained, and multi-stacked graphene layers, we show how to disentangle the different (partially counteracting) effects onto the position of the Raman lines.

## I. RESULTS AND DISCUSSION

### A. Raman response of stage II to stage VI GICs

We consider $KC_8$ as our starting potassium-graphite intercalation phase. Then we performed controlled temperature driven de-intercalation experiments to synthesize the higher intercalation stages: *stage II* ($KC_{24}$), *stage III* ($KC_{36}$), *stage IV* ($KC_{48}$), *stage V* ($KC_{60}$), and *stage VI* ($KC_{72}$). The corresponding Raman spectra at 568 nm laser excitation are depicted in Fig. 1 a). The G-line always displays a high frequency component $G_c$ between 1600 and 1610 cm$^{-1}$. This mode has a slight asymmetry due to the Fano interference of the conduction electrons in the electron doped charged layers and is related to the charged graphene layers next to an intercalant layer. This assignment is also confirmed by comparing to *stage II* $KC_{24}$ within the NL model, where each graphene layer is in contact with potassium atoms and just charged graphene layers exist, hence $KC_{24}$ exhibits only one G-line as $G_c$ at 1610 cm$^{-1}$. In addition, for stages higher than $KC_{24}$ a second line $G_{uc}$ appears around 1580 cm$^{-1}$. The position of the $G_{uc}$ is close to the G-line in pristine graphite (1583 cm$^{-1}$) and of graphene [20] (1580 cm$^{-1}$). Therefore $G_{uc}$ is assigned to the response of basically uncharged graphene layers surrounded by charged graphene layers. A detailed line shape analysis of the two G-line components using an asymmetric Fano line for the charged layers ($G_c$) next to an intercalant layer and a Lorentzian line for the uncharged carbon layers ($G_{uc}$) was performed. The results are closely matching the experimental spectra and are depicted as solid lines in the respective panels of Fig. 1 a) and Fig. 2. As shown in Fig. 1 b), the positions of these two components are not constant and have a linear dependence on the intercalation stage. This shift is not predicted by the NL model. We will explain the shift below in an analysis that takes into account the small fractional charging even of the inner layers and the lattice expansion due to intercalation.

We find a linear relation between the intensity ratio (R) of the the two G-line components ($G_{uc}$ and $G_c$), in agreement with the NL model. This relation is depicted by $R = I_{uc}/I_c$ in Fig. 1 c) for *stages III* to *VI*. Our results using the 568 nm laser excitation agree very well with the results of Solin et al. [17]. R is also independent from the type of alkali metal intercalant for the case of K, Rb, and Cs GICs [17]. In order to analyze these highly staged GICs in more detail we performed a complete multi-frequency analysis as depicted in Fig. 2 for five different laser excitations. A clear photon energy dependence of the intensity ratio between the charged and uncharged G-line components is obvious (see also Fig. 1 d), but the linear dependence of R is universal and can be safely used for the identification of the different stages (following the protocol explained below in the methods). This shows that in all cases the predominant charge transfer from the intercalant to the neighboring graphene layer is a valid approximation. The different slopes for different photon energies can be tentatively explained by a different frequency dependent resonance Raman cross section of $G_c$ and $G_{uc}$ and by different charge carrier absorption of the charged layers at high photon energies. This means that, in agreement with previous results, Raman spectroscopy can be used to unambiguously identify the different stages and correlate them to structural assignments by XRD.

We now turn to a detailed analysis of he double resonant 2D line. The NL model lets us expect that c and uc graphene layers give rise to separate 2D lines. We observe experimentally that the 2D-line is absent for *stage I* and *stage II* potassium GICs but present for the higher order stages and for pure graphene and graphite. This means that the 2D response is only due to the inner (uc) layers. The double resonance is suppressed for the outer (c) layers due to the strong charging. Therefore in *stage I* and *stage II* GICs where all graphene layers are adjacent to an intercalant, the 2D line is suppressed altogether. Interestingly this is not the case for p-type intercalation as recent results on $FeCl_3$ doped multilayer graphene, owing a doping level similar to Stage I GIC [14]. This different behavior might be due to a different localization of the charges transferred to the $FeCl_3$ ions upon p-type doping which should be investigated in future studies. For the potassium intercalation stages *III* to *VI*, the 2D line displays a clear dispersion with respect to the laser energy (right panels in Fig. 2) and possesses a fine structure that depends on the number of uc layers (see of Fig. 3 b).

Considering that the double resonant Raman peak arises from the inner graphene layers only, we can regard a the inner layer of $KC_{36}$ as a graphene mono-layer, the inner two layers of $KC_{48}$ as a bi-layer, the inner three layers of $KC_{60}$ as a tri-layer. In the analysis, we can thus draw an analogy with the splitting of the 2D-line in double- and few-layer graphene[1, 4]. A closer inspection of the 2D-line of the different GIC stages and fitting with two Voigtian functions, demonstrates indeed a splitting of this line starting with *stage IV* (see Fig. 3 b). We observe a splitting of 25.71, 38.34, and 39.48 cm$^{-1}$ between the two Voigtian functions of *stage IV*, *stage V*, and *stage VI* respectively, while *stage III* presents a 2D-line with one component only. In Fig. 3 a, we display the dispersion of the 2D line as a function of the laser energy. (For the stages where the 2D-line is split, we use the position of the lower peak which is much more intense than the upper peak as can be seen in panel (b) and as it is explained in Ref. [21].) The slope of the 2D dispersion as a function of the laser energy is about 99 cm$^{-1}$/eV for all the different stages. This agrees with the dispersion of the 2D line of natural graphite, turbostatic-graphite [9] and graphene/Si [9] which are also shown in Fig. 3 a.

However, the actual position of the 2D line depends on the stage of the GIC. E.g., the 2D line of the inner layer in $KC_{36}$ is down-shifted by about ∼36 cm$^{-1}$ as compared to the 2D line of pristine graphene. Similarly, the 2D lines of $KC_{48}$, $KC_{60}$, $KC_{72}$ are downshifted with respect to the ones of bi-, tri- and quad-layer graphene, respectively. The amount of the downshift decreases with higher stage number. In order to explore all possible reasons for this very pronounced down-shift, we have performed a detailed theoretical analysis based on *ab-initio* calculations of charge density, electronic dispersion, and phonon frequencies.

### B. Theoretical discussion of the Raman spectra

We discuss first the amount of charge transfer from the potassium intercalant layer to the different graphene layers. For this purpose, we have calculated the total charge density $\rho_{GIC}(z)$ as a function of the direction perpendicular to the layers, averaged over the in-plane directions (x-y-plane). From this, we subtract the "reference charge density", $\rho_{ref}(z) = \rho_C(z) + \rho_K$ which is the sum of the charge densities of the graphene layers and potassium layers calculated separately (in the geometry of the compound system). This method follows the earlier calculations by Hartwigsen et al. [22] on *stage I* intercalation compounds and by Ancilotto and Toigo [23] on potassium adsorbed on a graphite surface. The charge density difference, $\Delta\rho(z) = \rho_{GIC}(z) - \rho_{ref}(z)$, is thus a convenient quantity to visualize charge transfer: it is negative around the position of the K atoms which tend to donate their electrons and it is positive where those electrons are accumulated. Fig. 4 a) demonstrates this for *stage III* $KC_{36}$ under ABA-Stacking, and *stage VI* $KC_{72}$ under AAA-Stacking in Fig. 4 c) (approximated model). Obviously, the electrons donated by the potassium atoms accumulate mainly on the potassium boundary carbon layers. In order to calculate a value for the charge transfer, we define (somewhat arbitrarily), the limit between the potassium and the carbon layer (marked $i'$ in Fig. 4 a and c) as the value of $z$ where $\Delta\rho(z)$ changes sign. Integrating the density-difference curve between those limits, one obtains a charge transfer of 0.39 electrons from each potassium atom to the graphene layers. Table 1 contains detailed information on the charge accumulation per layer. Most of the transferred electrons accumulate on the outer graphene layer. The charge concentration on this layer is almost independent of the staging. In contrast, the charge concentration on the inner layers remains (relatively) low and varies with the stage number. This explains why both the $G_c$ and $G_{uc}$ components depend only weakly on the stage number. The splitting which was already observed by Solin and Caswell [17] can now be understood on the basis of the high (and constant) charge density of the charged graphene layers. Due to the breakdown of the Born-Oppenheimer expansion, charging of graphene leads to a strong stiffening of the G-mode [18]. By electrochemical top-gating, electron concentrations of up to $5 \times 10^{13}/cm^2$ have been achieved, [24] leading to a G-line position at about 1605 cm$^{-1}$. In GICs, the charge density on the $G_c$ component reaches similar values (see Table 1) which explains its high frequency. On the other hand the low frequency observed for the uncharged component $G_{uc}$ is also slightly affected by this charge transfer beyond the nearest neighbor. We obtained the G-line up-shift due to the charge density by using $\sigma$ from Table 1, and Eq.3 from Ref. [34] as it gives the response of the system under adiabatic+expanded lattice conditions. This yields a calculated $G_{uc}$ of between 1584 cm$^{-1}$ and 1586 cm$^{-1}$ for all different GICs. This is strongly overestimating the observed values and even yields the wrong trend, which highlights that there is some important ingredient missing in our explanation.

In order to address this point we first turn to the discussion of the 2D-line as a function of staging, making use of the double-resonance Raman model of Thomsen and Reich [25]. The model successfully describes the D and 2D dispersion as a function of laser energy as well as the splitting of the 2D line for double, triple, and multi-layer graphene [1, 4], provided that renormalization of the highest optical-phonon branch (HOB) due to electron-correlation effects is properly taken into account [26]. The different intercalation stages (2nd, 3rd, 4th, etc.) can be viewed, respectively, as bi-, tri-, and quadri-layer graphene, separated by K intercalant layers. One might thus expect a similar splitting and shifting of the 2D line as observed in Refs. [1, 4]. Instead, the experiments show an absence of the 2D-line for the *stage II* GIC, a single 2D-line for the *stage III* compound and an up-shift of the 2D line with increasing staging order (see Fig. 3). This difference can be understood as due to the charging of the graphene layers adjacent to the K atoms. In order to demonstrate this, we show in Fig. 4 b) the electronic band-structure (DFT-LDA) of *stage III* GIC $KC_{36}$ (in ABA-Stacking configuration). The unit-cell contains 24 atoms per carbon layer, thus the band-structure is plotted in a $2\sqrt{3}$ x $2\sqrt{3}$ supercell (compared to the primitive cell of graphene that contains only 2 atoms). In this supercell, the high-symmetry point K of the Brillouin zone of graphene is folded back onto $\Gamma$.

$KC_{36}$ displays notable exceptions from the linear crossing of the $\pi$-bands due to the interlayer interaction. Along with the band-structure of GIC, we plot the band-structure of pure graphene, displaying the linear crossing of the $\pi$-bands at $\Gamma$ (K in the primitive cell). We shift the graphene bands in energy such that they match the corresponding bands of $KC_{36}$. The red-dashed lines correspond to the electronic bands of the charged graphene layers. The Dirac point is shifted to $\Delta E_1 = 1.07$ eV below the Fermi level, corresponding to a strong charging. The green-dashed lines correspond to the electronic states of the weakly charged layer. Correspondingly, the Dirac point is shifted downwards only by $\Delta E_2 = 0.49$ eV. This shift of the Dirac point gives us an additional measure for the charge density of the layers: From the density of states of graphene, $n(E) = |E|/(\pi\hbar^2 v^2)$, where $v$ is the Fermi velocity of graphene, one obtains the charge density by integration over the energy from the Dirac point to the Fermi level: $\sigma(E_F) = ((E_F)^2 e)/(2\pi\hbar^2 v^2)$ This connection of Fermi-level shift and electron density was also used in Ref. [27] for the determination of the average doping level based on work-function measurements. With the DFT-LDA value of the Fermi velocity of $v = 0.85$x$10^6$ m/s [28], we obtain $\sigma_1 = 5.9$x$10^{13}$ cm$^{-2}$ for the charged layer and $\sigma_2 = 1.2$x$10^{13}$ cm$^{-2}$ for the uncharged layer in $KC_{36}$ (Fig. 4 a). The value of $\sigma_1$ is 13% larger than the corresponding value in Table 1, while the value off $\sigma_2$ is 47% smaller than the value in Table 1. These differences give an indication of the uncertainties of different charge transfer assignments (the electrons localized in between layers cannot be unambiguously assigned as belonging to one or the other layer).

Concerning the 2D-line results of $KC_{36}$, we need to answer two questions: (i) why is the 2D-line not split into two peaks like the G-line? (ii) Why is the 2D line down-shifted by about 40 cm$^{-1}$ compared to pure graphene?

In Fig. 4 b), vertical arrows mark dipole-allowed electron-hole pair transitions corresponding to a laser energy of 2.3 eV. Since in this energy range, the $\pi$-bands of $KC_{36}$ almost exactly match the (shifted) $\pi$-bands of pure graphene, the transitions take place at the same electronic wave-vector. In the double-resonance Raman model the electron/hole performs a quasi-horizontal transition to a state in the vicinity of the neighboring point K'. This means that phonons with equal wave-vector $\mathbf{q}$ are excited in graphene and $KC_{36}$. The red vertical arrow marks a transition where the excited electron is barely above the Fermi level. This transition (in the charged layer) is therefore strongly suppressed with respect to the one marked by the green vertical arrow which is an electronic excitation in the uncharged layer. Thus, contrary to the G-line, only one component of the 2D-line is present in the spectrum of $KC_{36}$ (and no 2D-line is visible for *stage II* GIC $KC_{24}$ where only the strongly charged layers exist).

We have also considered the blue vertical transition from the $\pi$-band of the charged layer to the $\pi^*$ band of the uncharged layer. However, since it is an inter-layer transition, its oscillator strength is negligible compared to the one of the intra-layer excitations. Since the electronic structure of the uncharged layer is decoupled from the one of the charged layer (as manifested by the almost rigid band shift in Fig. 4 b), one might expect little or no difference in frequency between the 2D line of isolated graphene and the one of $KC_{36}$. Indeed, several reasons could be found (related to the strong Kohn anomaly of the highest-optical branch at K [29]) that could even explain a slight up-shift of the 2D-line: (i) The residual charging of the inner layer leads to a reduction of the electron-phonon coupling around K [30] and might increase the 2D position by 10 cm$^{-1}$ (extrapolated from Fig. 2d of Ref. [12]). (ii) The opening of a gap between the $\pi$ bands at $\Gamma$ (K) could lead to a partial suppression of the Kohn-Anomaly in analogy to what happens for graphene in close contact to a Ni(111) surface [31]. (iii) The dielectric screening by the quasi-metallic environment could reduce the Kohn-Anomaly as recently observed for graphene on dielectric substrates. In order to check if any of the above three arguments holds, we performed a phonon calculation [32] of the HOB at K, comparing the mode of the single layer with the mode of the inner layer of *stage III* potassium GIC. The mode of the uncharged layer has a frequency of 1269.6 cm$^{-1}$ while the mode of the single layer has 1271.9 cm$^{-1}$. We conclude that the phonons of the HOB around K remain essentially unchanged if the lattice constant of KC$_{36}$ is the same as the one of graphene. However, there are subtle changes for the lattice constant as a function of staging. Nixon and Parry have measured the expansion of the carbon-carbon bond length in potassium GIC (see Table 4 in Ref. [15]). The lattice constant of KC$_{36}$ is 0.20% larger than the one of graphite. In order to verify this experimental result, we have performed a full cell-optimization of KC$_{36}$ and of pure graphite: we obtain a theoretical bond-length expansion of ∼0.22% in very good agreement with the experiments.

Therefore we can use the experimental lattice expansion from the XRD measurements in Ref. [15] of different GICs (whose values are also given as last column in Table 1) to evaluate the redshift of the Raman response. This is equivalent to putting strain on the individual graphene layers. This strain induced lattice expansion in graphene has previously been related to the phonon frequency of the G and 2D line using the Grüneisen parameter [2, 7] in experiments applying uni-axial [4, 6, 8] and bi-axial strain[7]. In our experiments, the regular incorporation of potassium atoms in between the layered structure of graphite, results in bi-axial strained graphene layers. The shift of Raman frequency as function of strain was calculated by using the relation $\gamma = -1/\omega_0 \cdot \partial\omega/\partial\varepsilon$ where $\omega_0$ is the Raman frequency without strain, $\partial\varepsilon$ is our calculated bi-axial strain in GIC, and $\gamma$= 2.2, and 3.3 is the Grüneisen parameter for the G and 2D band phonons respectively [7].

For KC$_{36}$ we have calculated the phonon-frequency shift of pristine graphene, and lattice expanded graphene. For the HOB between K and M the frequency down-shift is about 18 cm$^{-1}$. This corresponds to a 2D-line redshift by 36 cm$^{-1}$. Thus, we can conclude that the redshift of the 2D-line (with respect to graphene) is almost entirely due to the small (but non-negligible) lattice expansion of the GIC. This lattice expansion also has a profound influence on the G-line position. This latter is, however, also influenced by the charge transfer as will be described in detail below.

## C. Charge transfer and bi-axially strained graphene layers in GIC

These two factors had been linked by Pietronero and Strässler [33] considering the well established C-C bond length in GIC as a main tool to determine the charge transfer in those systems. Hitherto, these two main factors directly affect the Raman response in GIC by: (i) the induced charge transferred from the K atoms into the carbon layers, and (ii) the in-plane strain coming from the change in the C-C bond length. The results from this analysis are depicted in Fig. 5.

In the left panel (Fig 5a) we show the frequency dependence of the G-line components as a function of inverse stage. Experimentally, for both the $G_{uc}$ and the $G_c$ component (red circles), we observe a linear decrease in frequency from *stage VI* to *III*. The slope as function of inverse stage (red dashed line) is slightly lower for $G_{uc}$. In addition, the C-C bond length of the different GIC from XRD studies of the in-plane lattice expansion of Ref. [15] are shown on the top axis. In order to understand the staging dependence of both G-line components, we have to add the effect of lattice expansion and of charge transfer. Starting from the G-line position of pristine graphene [1, 20] (black dashed line) we have to add the upshift from the increased charge density on the layers and the down shift from the bi-axial strain on the graphene layers and the corresponding change in the C-C bond length. We show these two contributions to the G-line as vertical green arrows (for the charge transfer related stiffening including the corresponding lattice expansion [34]) and vertical blue arrows (for the additional effective bi-axial strain), respectively. For the $G_{uc}$ component we can evaluate the upshift due to the increased charge density using our calculated charge transfer to the $G_{uc}$ summarized in Table 1 (green open circles in Fig. 5a). The resulting charge transfer to the $G_{uc}$ is lowest for $KC_{72}$ and has an approximately linear increase with inverse stage. Concomitantly, the downshift due to the effective biaxial strain (blue open circles) are also shown. The resulting frequencies of the $G_{uc}$ component in GIC (blue crosses) perfectly match the values of the experimental $G_{uc}$. In order to confirm the additivity of the effects of charge transfer and lattice expansion, we also performed non-adiabatic calculations[41] of the phonon frequency of charged layers a fixed lattice constant. For $KC_{36}$ this yields a nominal upshift of 13.8 cm$^{-1}$ for the $G_{uc}$, which together with the downshift of 15 cm$^{-1}$ (determined from the experimental total lattice expansion determined by XRD using the Grü neisen parameter for the G phonons of Ref. [7]), yields a position of 1579.6 cm$^{-1}$. This is in excellent agreement with the results using an effective strain. We also verified computationally (by variation of the lattice constant for neutral and charged graphene) that the Grüneisen parameter remains constant for the charge values observed in GICs. This means that both effects are truly additive. For the G$_{uc}$, about 50% of the observed experimental average lattice expansion measured by XRD are related to the charge transfer from the intercalants.

For the highly doped G$_c$ component the story is more complex. Experimentally the Raman response of highly doped graphene has been achieved by polymer electrolyte gating [24, 35] and by alkali metal vapor dosing [36]. Interestingly for a broad range of electron concentrations between 4·10$^{13}$/cm$^{-2}$ and 10·10$^{13}$/cm$^{-2}$ a position of 1611 cm$^{-1}$, very similar to the 1610 cm$^{-1}$ mode in highly charged KC$_{24}$, is observed [36]. This saturation in the Raman frequency at high doping is also reported theoretically [34], also the absolute frequency is not correctly addressed in the theoretical description at high charge transfer. Therefore, in a broad doping range the charge transfer induced strain compensates the non-adiabatic effects. For this reason, we used the experimental position of the "pristine" G$_c$ since we observe for all intercalation compounds a charge transfer to the charged layers above 4·10$^{13}$/cm$^{-2}$ (see Table 1). Interestingly, although one would expect a higher contribution of the charge transfer to the effective strain, adding the the same effective bi-axial strain as for the G$_{uc}$ we find a nearly perfect match of the resulting frequency (blue crosses) with the experimental G$_c$. This confirms that the G-line response in GIC is related to charged and strained graphene layers. We point out that this model cannot be directly applied for stage I and stage II GIC because there are no uncharged graphene layers, the charge transfer is much stronger and the line position is influenced by the Fano interference with the conduction electrons [17, 37].

The same analysis performed to the second order 2D line (strongest component related to the main second order Raman process) is shown in Fig. 5b. We observe a linear decrease in position of the main 2D line as function of the inverse stage (red circles). The vertical dashed blue lines again correspond to the strain induced downshift of pristine graphene. The blue labels in the figure correspond to the relative percentage of bi-axial strain from [15] and the evaluated downshift using the Grüneisen parameter for the 2D-line [7]. Due to the higher frequency of the 2D line a much bigger shift between 10 and 40 cm$^{-1}$ is observed. The upshift due to the residual weak charging can be estimated to be less than 2 cm$^{-1}$ [4] and can be neglected. Therefore, we are able to artificially "release" the biaxial strain and compare the 2D-line in "unstrained "$KC_{36}$, $KC_{48}$, $KC_{60}$, and $KC_{72}$ (blue crosses in Fig. 5b) to the experimental position of unstrained mono-, bi-, tri-, and quad-layer graphene from Ref. [4] (black squares in Fig. 5b). We find a very good agreement. This confirms that the 2D-line in GIC comes from strained multi-layer graphene.

In addition we can go one step further and compare the 2D line position to absolute in-plane lattice constants of the graphene layers based on the XRD results. This relies on the fact that Nixon and Perry [15] observed a linear relation of the in plane lattice constant versus inverse stage of the GIC. This allows to use the measured lattice constants to accurately correlate the 2D frequency to the C-C bond length shown as the right y-axes of 5 b). Thus, one can use the 2D Raman response directly to determine the in-plane lattice constant of mono- and multi-layer graphene as a function of their internal strain even on an absolute scale. This demonstrates that Raman spectroscopy is a very powerful tool to identify local internal strain in single and few-layer graphene and their nanoelectronic devices and composites, yielding even absolute in plane lattice constants.

## II. CONCLUSIONS

In summary, we have analyzed the intrinsic Raman response from strained graphene layers in graphite intercalation compounds. For *stage III* and higher there are two nearest layer environments: heavily charged graphene layers adjacent to an intercalant layer and basically uncharged graphene layers sandwiched between other graphene layers. By *ab-initio* calculations of the charge densities and the electronic band dispersions, we have demonstrated that the charge transfer is incomplete (less than 1 electron per potassium atom) and that most (but not all) of the transferred charge remains on the charged graphene layers adjacent to the intercalants. This leads to an electronic decoupling of the inner (uncharged) from the outer (charged) layers and consequently also to a decoupling of the corresponding Raman spectra: The G-line splits into two peaks and the 2D line is entirely due to the uncharged inner layers while the 2D line of the outer layers is suppressed due to the strong charging. The quantitative interpretation of the peak positions requires that the internal strain of the graphene layers is taken into account. This allows to unambiguously identify the Raman response of strained charged and uncharged graphene layers and to correlate it to the in-plane lattice constant determined by XRD. Therefore Raman spectroscopy is a very powerful tool to identify local internal strain in uncharged and weakly charged single and few-layer graphene yielding even absolute lattice constants. This has important implications for the application of Raman spectroscopy to identify for instance the strain in nanocarbon based nanoelectronic and optoelectronic devices as well as the local interfacial strain in graphene and carbon nanotube polymer composites on an absolute scale.

## III. MATERIALS AND METHODS

The intercalation experiments on natural graphite single crystals have been conducted *in-situ* in a vacuum better than $\sim 4 \times 10^{-8}$ mbar keeping the graphite sample inside a quartz ampoule with a flat surface. The Raman measurements have been performed using 458, 488, 514, 568, and 647 nm excitation wavelengths at 1.2 mW between 1400 cm$^{-1}$ to 3000 cm$^{-1}$. Potassium with a 99.95% purity (Aldrich) was evaporated until bright golden graphite crystals which are assigned to stage I KC$_8$ were obtained [37]. Subsequently, the sample was resistively heated to 200°C until it turned homogeneously blue characteristic color of *stage II* KC$_{24}$). A controlled *in-situ* high vacuum high temperature de-intercalation process was performed by increasing the temperature in steps of 50°C [13]. The line positions were corrected by employing calibration lamps. Six potassium GIC *stage I* to *VI* (KC$_8$, KC$_{24}$, KC$_{36}$, KC$_{48}$, KC$_{60}$, and KC$_{72}$) were identified for all laser lines used. Once each stage was identified we followed a protocol acquiring multi-frequency Raman spectra keeping constant the acquisition region, and maintaining the lowest possible exposure time to avoid laser induced de-intercalation as for instance reported by [38] and later on by us [37].

The calculations were performed using density-functional theory in the local density approximation. We have used the code **quantum-espresso** [39]. For the values of the K-C and C-C inter-plane distances as well as for the in-plane lattice constants, we used the experimental values given in Table 4 of Ref. [13], i.e., $a = 1.42 \mathring{A}$ for the bond-length, $d = 3.35 \mathring{A}$ for the inter-plane distance in graphite, and $d_M = 5.41 \mathring{A}$ for the distance between the intercalated layers. For simplicity, in order to keep super-cell size low, we used AA-stacking in adjacent graphene-layers, except for KC$_{36}$ where we used ABA-stacking. K atoms were placed between the centers of carbon hexagons. No geometry relaxation was undertaken. The reciprocal unit-cell was sampled by a 6×6×2 Monckhorst-Pack grid. Norm-conserving pseudopotentials (with nonlinear core correction for K) and an energy cutoff at 60Ry were used.

**ACKNOWLEDGMENTS**

We acknowledge for the financial support of the project FWF-I377-N16, the OEAD AMADEUS PROGRAM financing. L.W. acknowledges funding by the ANR (French National Research Agency) through project ANR-09-BLAN-0421-01. Calculations were done at the IDRIS supercomputing center, Orsay (Proj. No. 091827), and at the Tirant Supercomputer of the University of Valencia (group vlc44).

TABLE I. Calculated charge transfer (e$^-$ per K atoms) from the intercalated K atoms to the graphene-layers for *stage III* to *VI* potassium GIC. The last column gives the bi-axial strain of the graphene layers [15].

| | el. per K atom | | | | $\sigma$ ($10^{13}$/cm$^2$) [a] | | | Bi-axial strain (%) |
|---|---|---|---|---|---|---|---|---|
| | K | 1st | 2nd | 3rd | 1st | 2nd | 3rd | |
| KC$_{36}$ | -0.39 | 0.33 | 0.12 | - | 5.2 | 1.9 | - | 0.20 |
| KC$_{48}$ | -0.39 | 0.28 | 0.11 | - | 4.5 | 1.7 | - | 0.13 |
| KC$_{60}$ | -0.39 | 0.26 | 0.11 | 0.04 | 4.1 | 1.7 | 0.6 | 0.10 |
| KC$_{72}$ | -0.39 | 0.26 | 0.10 | 0.03 | 4.1 | 1.6 | 0.5 | 0.06 |

[a] The corresponding electron density is given in electrons/cm$^2$.

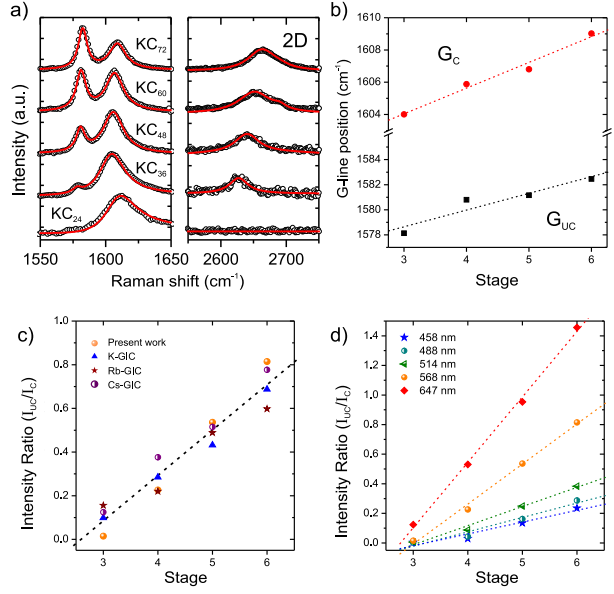

FIG. 1. Raman spectra of *stage II* to *VI* GIC measured with a laser wavelength of 568 nm. In panel a) the spectra (black dots) together with the results of a line shape analysis (red line) in the region of the G and 2D line are depicted. Panel b) shows the positions of the two G-line components ($G_c$ and $G_{uc}$ as a function of stage $n$. Panel c) and panel d) shows the intensity ratio (R) of the two components as function of $n$. In panel c) R was calculated from the Raman intensity of the low ($G_{uc}$) and high frequency ($G_c$) modes in the G-line and compared to the literature values [17] for K, Rb and Cs intercalation measured with a 514 nm wave length. Panel d) shows the photon energy dependence of R together with a linear fit of R=$I_{uc}/I_c$ (dashed line).

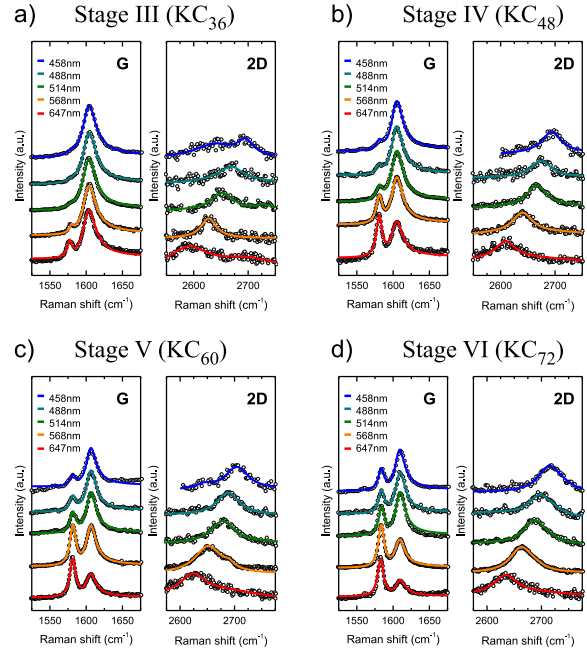

FIG. 2. Raman spectra from *stage III* to *stage VI* potassium GIC. The different spectra were acquired using laser-excitation wavelengths in the range of (458 nm and 647 nm). In the left panel of a), b), c) and d) the G-line of each intercalation compound is shown; in the right panels of each stage the 2D-line is depicted. The solid lines in the figure are fits using Voigthian and Fano line shapes (see text), respectively.

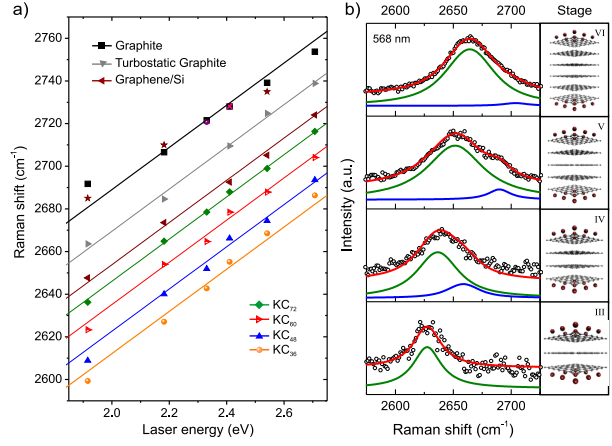

FIG. 3. a) Dispersion of the most intense double-resonant 2D Raman mode in GIC, graphite and graphenes as function of laser energy. The different symbols depict the experimental 2D-line position measured. The values of graphene/Si, Turbostatic-graphite, and pristine graphite were extracted from Ref. [9], and [4, 10, 40] respectively. b) Detailed line-shape analysis of the 2D-line in *stage III to VI* is shown using up to two Voigthians (green and blue lines). The result of the analysis is shown as red line.

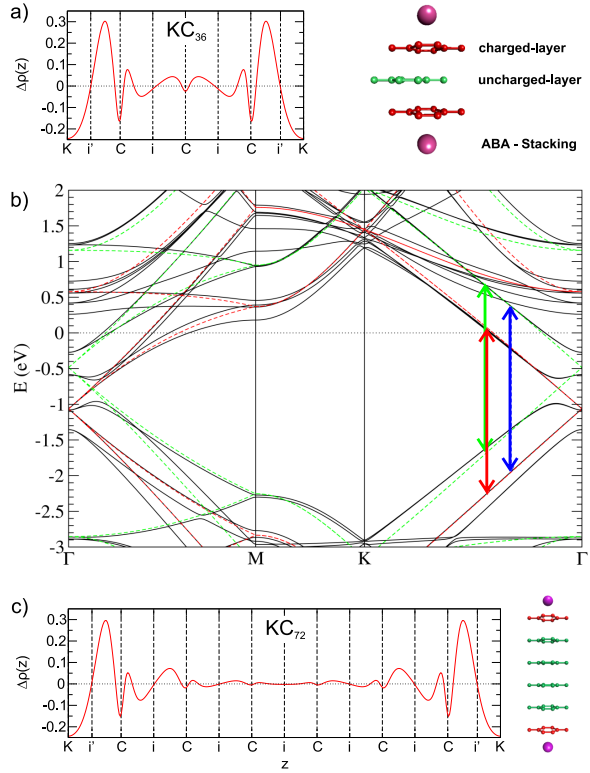

FIG. 4. Charge density and Bad structure analysis. a) Charge density distribution for $KC_{36}$ with a sketch of the charged and uncharged layers. In the charge density analysis, the position of planes consisting of carbon/potassium atoms is marked by C/K, respectively. Mid-points between graphene planes are marked by $i$ and separation of the K and C planes, defined as the position where the density difference crosses 0, is marked by $i'$. In panel b), the band-structure of $KC_{36}$ (black solid lines) and of pure graphene from our *ab-initio* is shown. The pristine graphene bands are shifted in energy to match the bands of the charged layer (red-dashed lines) and of the uncharged layer (green dashed lines). The red, green and blue vertical arrows mark the transition between the $\pi$-bands of the charged-charged, uncharged-uncharged and charged-uncharged layers at 2.3 eV laser energy. In panel c) the same analysis of the charge distribution as in a) is depicted for $KC_{72}$.

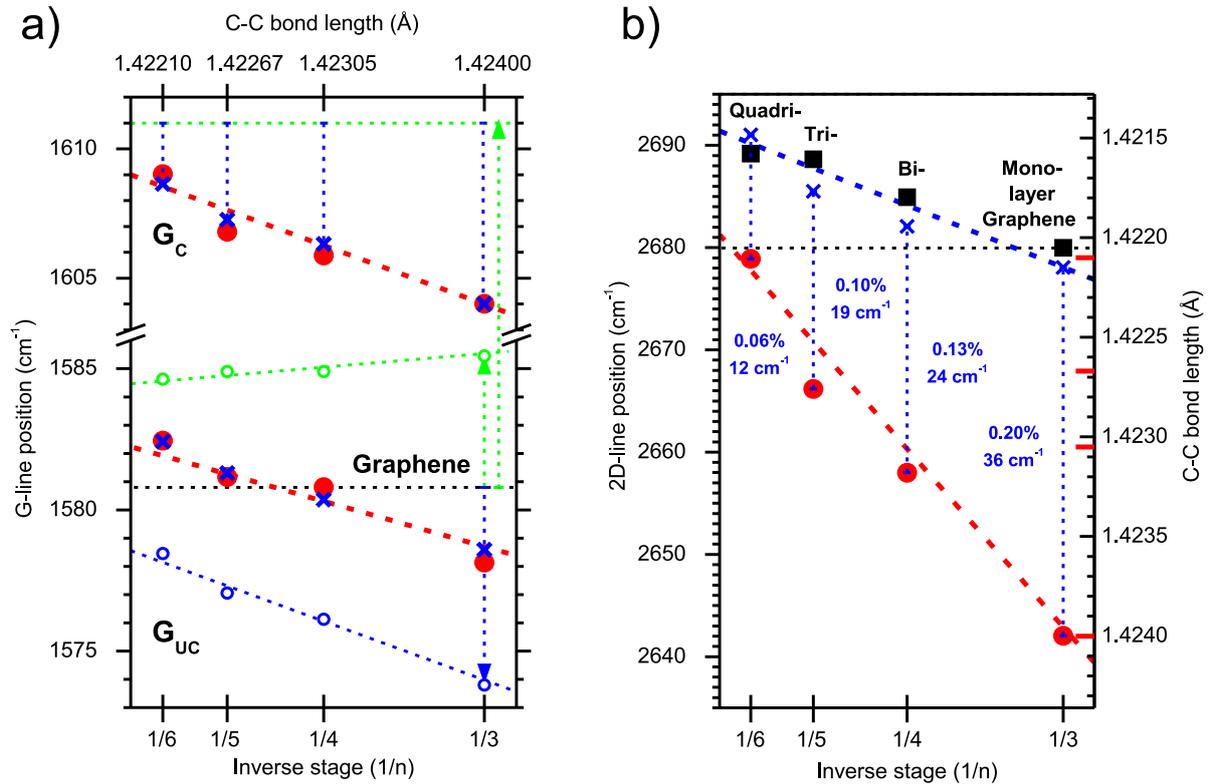

FIG. 5. In panel a) the G-lines of the potassium GIC is depicted as function of the inverse stage. The upper x-scale depicts the C-C bond length of the XRD results of Ref. [15]. The dashed green lines and the green open circles show the G-line upshift due to the high charge transfer to the charged $G_c$ component and the remaining small charge transfer to the $G_{uc}$ component. The blue dashed line and blue open circles come from the bi-axial strain induced softening of the G-line of graphene (black dashed line). The blue crosses depict the positions after adding the charge transfer and subtracting the internal strain. In panel b) the 2D-line position of the high-frequency-mode of GIC (red circles) and of unstrained (multi-layer)graphenes from Ref. [4] (black squares) is plotted as function of inverse stage. The dashed blue lines values in the figure depict the frequency softening by biaxial strain. The blue crosses depict the positions after subtracting this internal strain. The second y-scale depicts the C-C bond length owing a linear relation to the 2D line. The short red line are the experimental XRD bond length of the upper x-axes in panel a).